\documentclass[%
 reprint,
 amsmath,amssymb,
 longbibliography,
 aps,
]{revtex4-1}

\usepackage[english]{babel}
\usepackage[normalem]{ulem}
\usepackage{graphicx}
\usepackage{dcolumn}
\usepackage{bm}
\usepackage{xcolor}
\usepackage{tensor}
\usepackage{comment}

\newcommand{\eps}{\varepsilon}
\DeclareMathOperator{\Tr}{Tr}
\begin{document}


\title{Dionysian Hard Sphere Packings are Mechanically Stable at Vanishingly Low Densities}

\author{R. C. Dennis}
\author{E. I. Corwin}
\affiliation{Department of Physics and Materials Science Institute, University of Oregon, Eugene, Oregon 97403, USA.}

\date{\today}

\begin{abstract}
High strength-to-weight ratio materials can be constructed by either maximizing strength or minimizing weight. Tensegrity structures and aerogels take very different paths to achieving high strength-to-weight ratios but both rely on internal tensile forces. In the absence of tensile forces, removing material eventually destabilizes a structure. Attempts to maximize the strength-to-weight ratio with purely repulsive spheres have proceeded by removing spheres from already stable crystalline structures. This results in a modestly low density and a strength-to-weight ratio much worse than can be achieved with tensile materials. Here, we demonstrate the existence of a packing of hard spheres that has asymptotically zero density and yet maintains finite strength, thus achieving an unbounded strength-to-weight ratio. This construction, which we term Dionysian, is the diametric opposite to the Apollonian sphere packing which completely and stably fills space. We create tools to evaluate the stability and strength of compressive sphere packings. Using these we find that our structures have asymptotically finite bulk and shear moduli and are linearly resistant to every applied deformation, both internal and external. By demonstrating that there is no lower bound on the density of stable structures, this work allows for the construction of arbitrarily lightweight high-strength materials.
\end{abstract}

\maketitle

%
When sand is densely packed, it is strong enough to support the weight of an elephant. But how loosely can one pack sand before this rigidity is lost? The answer is as loosely as one would like. That is, it is possible to rigidly pack hard spheres at any density, from filling all of space to filling none. In this manuscript we show a method for creating the sparsest possible hard sphere packings and demonstrate their impressive stability. Hard sphere packings are of particular interest because unlike other materials with a high strength-to-weight ratio such as tensegrity structures~\cite{fuller_synergetics_1982} and aerogels~\cite{kistler_coherent_1931}, hard spheres are purely compressive and do not rely on internal tensile forces.

There exist mechanically rigid packings with a density arbitrarily close to unity, such as the Apollonian gasket~\cite{bourke_introduction_2006, lagarias_beyond_2008}. We wish to find the foil to such a packing, that is, one with the smallest possible packing fraction that remains mechanically stable. As Dionysus is the nadir to the zenith that is Apollo~\cite{del_caro_dionysian_1989}, we refer to the sparsest possible mechanically stable packings as \textit{Dionysian packings}. We present in this manuscript a construction for a Dionysian packing which has vanishingly low density in two and three dimensions.

Rigidity~\cite{connelly_rigidity_2008} describes a state in which no motion is possible. In the context of sphere packings, this is termed \textit{strictly jammed}~\cite{torquato_breakdown_2003,donev_jamming_2004, donev_linear_2004, torquato_toward_2007, torquato_jammed_2010}. A strictly jammed packing is resistant to all possible volume preserving deformations of the particles and boundaries. 

Demonstrating that a packing is mechanically stable is commonly done using a linear programming algorithm~\cite{donev_jamming_2004, donev_linear_2004, torquato_toward_2007}. In addition to demonstrating that our packings are stable through this same linear programming approach, we also compute the elastic moduli for the underlying spring network.

Finding a Dionysian packing is the same as finding the jamming threshold of sphere packings~\cite{torquato_toward_2007, torquato_jammed_2010}. The jamming threshold is the lowest density that can be achieved for strictly jammed configurations. However, while this threshold has mostly been explored for monodisperse configurations, we show that lower density packings can be found by expanding the search space to include polydispersity. 

The method we employ is inspired by the construction of the B{\"o}r{\"o}czky bridge packing~\cite{boroczky_k_uber_1964, kahle_sparse_2012} for which locally stable bridges of circles can be constructed with arbitrary length. These bridges lead to packings with asymptotically zero density, but only satisfy the very weakest definition of stability; they are only \textit{locally stable} or \textit{locally jammed}~\cite{boroczky_k_uber_1964, torquato_breakdown_2003,donev_jamming_2004, donev_linear_2004, torquato_toward_2007, torquato_jammed_2010, kahle_sparse_2012}. Following the spirit of the B{\"o}r{\"o}czky bridge packing and allowing for the radii of the spheres to be additional degrees of freedom, we achieve Dionysian packings subject to periodic boundary conditions at arbitrarily low densities. This demonstrates that the lower density bound for mechanically stable, repulsive circle and sphere packings is precisely zero.

To determine if a packing is strictly jammed, we model it as a spring network in which spheres interact through a harmonic contact potential in their overlaps. We examine whether or not the spring network represents a minimum with respect to position degrees of freedom, $x,$ as well as symmetric affine, volume-preserving strain degrees of freedom, $\eps$~\cite{donev_energy-efficient_2003, donev_linear_2004} where the potential is
\begin{eqnarray}
U=\frac{1}{4}\sum_{i}\sum_{j\neq i}\xi_{ij}^2
\end{eqnarray}
and $\xi_{ij}$ is the normalized overlap between spheres $i$ and $j.$

We require force balance on all degrees of freedom.  The forces on the position degrees of freedom are
\begin{eqnarray}
F_{i}^{\alpha}=-\frac{\partial U}{\partial x_{i}^{\alpha}}=\sum_{k\in\partial i}\left(\frac{\xi_{ik}n_{ik}^{\alpha}}{r_i+r_k}\right)=0
\end{eqnarray}
where $n_{ik}^{\alpha}$ is the $\alpha$-component of the normalized contact vector pointing from particle $k$ to particle $i$ and $r_i$ is the radius of sphere $i.$  Forces on the strain degrees of freedom are
\begin{align}
-\frac{\partial U}{\partial \eps^{\alpha\beta}}=
\frac{1}{4}\sum_{i}\sum_{j\in\partial i}\frac{\xi_{ij}}{r_i+r_j}\left(n_{ij}^{\alpha}x_{ij}^{\beta}+n_{ij}^{\beta}x_{ij}^{\alpha}\right)
\end{align}
for spheres $i$ and $j$ in Cartesian directions $\alpha$ and $\beta$ where $\tensor{\eps}{^\alpha^\beta}$ is the strain degree of freedom and $x_{ij}^{\alpha}$ is the contact vector which is not normalized. 

These forces are subject to the volume-preserving constraint, $\Tr{\left(\eps\right)}=0$~\cite{donev_linear_2004} so that force balance is achieved when
\begin{align}
-\left.\frac{\partial U}{\partial \eps^{\alpha\beta}}\right|_{\Tr{\left(\eps\right)}=0}=-\frac{\partial U}{\partial \eps^{\alpha\beta}}+\frac{\delta^{\alpha\beta}}{d}\sum_{\gamma=1}^{d}\frac{\partial U}{\partial \eps^{\gamma\gamma}}=0.
\end{align}
Because this derivative is proportional to overlap, it is trivially zero for any packing where overlaps do not occur. To ensure that these packings are at a critical point due to a balancing of strain degrees of freedom, we evaluate the derivative with infinitesimal overlap.

The rigidity matrix~\cite{connelly_rigidity_2014} in conjunction with a linear programming algorithm~\cite{donev_jamming_2004, donev_linear_2004, torquato_toward_2007} is used to determine if packings are strictly jammed. The rigidity matrix, $R_x,$ relates a perturbation of the particles, $\vec{x},$ with the stresses on the bonds, $\vec{b},$ such that $\vec{b}=R_x\vec{x}.$ However, perturbing the particles is not our only degree of freedom to explore when considering whether or not a packing is strictly jammed as we must also consider bulk deformations of the system as encoded in strain degrees of freedom. We define the extended rigidity matrix as $R=
\begin{pmatrix}
R_x & R_{\eps}
\end{pmatrix}$
where $R_x$ is the ordinary rigidity matrix and $R_\eps$ relates the bond stresses to the strain degrees of freedom. (See supplementary materials for more information.) However, applying a strain that increases the volume of the periodic cell will allow all of the bonds to break, unjamming the packing. As such, we apply a constraint preventing the strain matrix, $\eps,$ from having volume changing deformations~\cite{donev_linear_2004}.

We quantify the degree of stability by calculating the resistance of the packing to compressive deformations and shear deformations via the bulk and shear moduli respectively. These quantities can be calculated simultaneously by computing the stiffness matrix, $C,$~\cite{liao_stress-strain_1997} for the packing. This matrix has the property $\vec{\sigma}=C\vec{\epsilon}$ where $\vec{\sigma}$ is the stress experienced by the packing when a particular strain, $\vec{\epsilon},$ is applied. The stiffness matrix can be computed in terms of the rigidity matrix as well as the states of self stress for $R_x.$ The matrix of states of self stress, $S,$ is an orthonormal basis for the zero modes of $R_x^T$ such that $R_x^T S=\vec{0}.$ The states of self stress represent the basis of stresses that can be placed on the bonds without causing particle perturbations. Using these terms, the stiffness matrix can be computed as
\begin{eqnarray}
C=R_\eps^T SS^T R_\eps.
\end{eqnarray}
(See supplementary materials for a derivation and an explanation of this equation.) 

To explicitly satisfy the constraints for shear stability and jamming, we focus on creating a packing which is locally stable and has a high number of contacts per particle, $z,$ and then test for stability. As illustrated in Figure~\ref{fig:construction} and described in more detail in the supplementary materials, this is achieved by placing $n$ circles labeled $a,$ where $n$ is an odd integer greater than 2, on a strictly convex curve $\mathcal{C}$ such that they kiss their neighbors. A new row of circles, $b,$ are then placed below so that each $b$ circle kisses two $a$ neighbors from below and a $b$ neighbor on each side. Finally, the centers of circles $c$ are placed on a line of zero slope and constrained to touch two $b$ circles from below. Applying the appropriate symmetries, a stable bridge is formed. This construction can be replicated and the bridges can be joined such that a circle packing is formed without overlapping regions. This packing, with the addition of thirteen circles filling the largest void, is a Dionysian packing for particular construction parameters. Our bridge placement for the two dimensional Dionysian packing is based on the contact network of the triangular lattice.

\begin{figure*}[]
\includegraphics[width=1\textwidth]{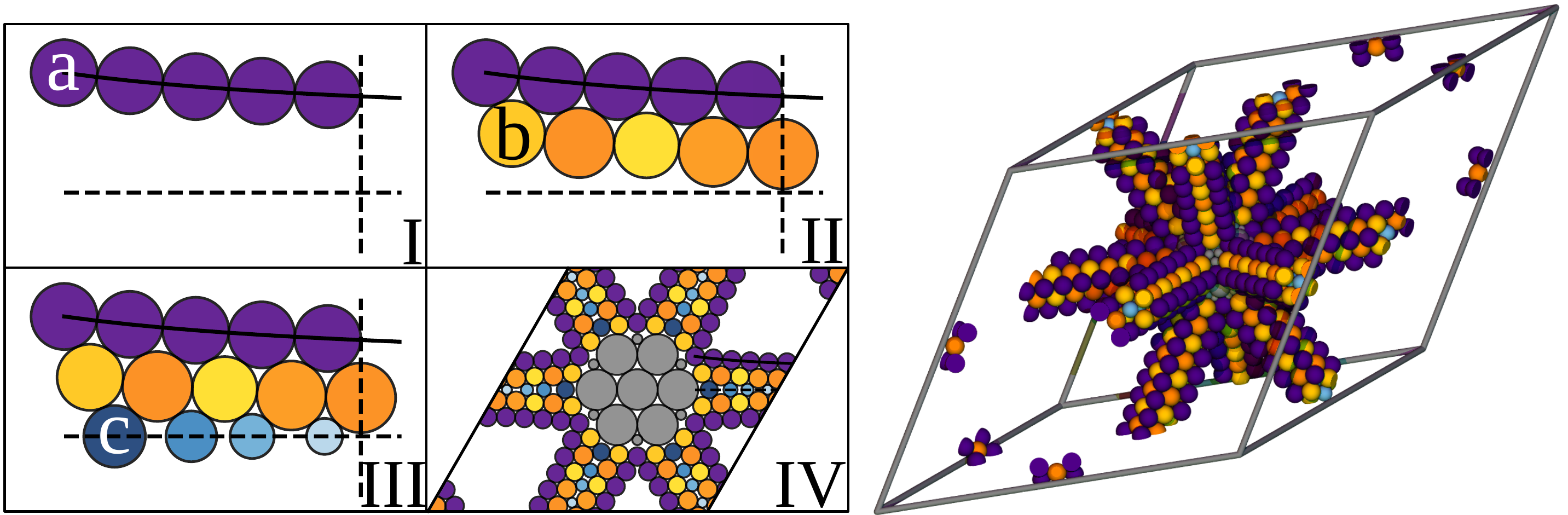}
\caption{The construction of a Dionysian packing in two and three dimensions. \textbf{Left.} \textbf{I)} A row of $n=5$ circles $a$ (purple) lie on a strictly convex curve $\mathcal{C}$ such that each circle kisses its neighbors. \textbf{II)} A row of $n=5$ circles $b$ (orange/yellow) are placed such that they kiss two circles $a$ from below and a circle $b$ on either side. The rightmost $b$ circle is constrained such that its center lies on the vertical line tangent to the rightmost $a$ circle. \textbf{III)} A row of $n-1=4$ circles $c$ (blue) lie on a horizontal line and kiss two $b$ circles above. \textbf{IV)} A bridge is formed by reflecting the circles about the dotted lines of symmetry. Three bridges are combined and their centers are filled as shown (gray). The resulting packing, which is jammed and shear stable, has a very low density and is a Dionysian packing in the limit as $n\to \infty.$
\newline
\textbf{Right.} A three dimensional mechanically stable packing at arbitrarily low densities. Such a construction contains the same three types of spheres as in the two dimensional analog but with additional symmetries and an entirely unrelated set of spheres filling the void region (gray). The three dimensional Dionysian packing has a much narrower set of convex curves $\mathcal{C}$ for which overlaps do not occur (as detailed in the supplementary materials). This requires a much more subtle curvature of $\mathcal{C}$ which is not apparent to the naked eye in this figure.}\label{fig:construction}
\end{figure*}

In the limit of an infinitely large bridge, we find that every $a$ circle has four contacts, every $b$ has six, and every $c$ has four. The asymptotic number ratio of this packing is $a:b:c=2:2:1.$ This means that there are $z=\left(2\times4+2\times6+4\right)/5=4\frac{4}{5}$ contacts per particle in two dimensions, which is larger than is required by the Maxwell rule for shear stable and jammed systems~\cite{lubensky_phonons_2015}.

For the B{\"o}r{\"o}czky locally jammed packing~\cite{boroczky_k_uber_1964, kahle_sparse_2012}, the two dimensional version can be used to create a locally jammed packing in any dimension by elevating the circles to spheres of the desired dimension and stacking the result. Such a trivial procedure will not work to extend the Dionysian construction because it results in structures which are not convex and so are subject to zero energy modes. To create a three dimensional Dionysian packing, we instead construct a set of six bridges in three dimensions and combine them as shown in Figure~\ref{fig:construction}. A three dimensional bridge is constructed very similarly to the two dimensional bridge and exploits the symmetries of three dimensional space. 

In the limit of an infinitely large bridge, we find that every $a$ sphere has six contacts, every $b$ has eight, and every $c$ has eight. The asymptotic number ratio for these spheres is $a:b:c=4:4:1.$ This means that there are $z=\left(4\times6+4\times8+8\right)/9=7\frac{1}{9}$ contacts per particle in three dimensions, which is larger than is required by the Maxwell rule for shear stable and jammed systems~\cite{lubensky_phonons_2015}.

Not all convex curves $\mathcal{C}$ result in viable packings; some choices of $\mathcal{C}$ result in overlapping of spheres in the limit as $n$ approaches infinity. While infinitely many viable choices of $\mathcal{C}$ are possible, for simplicity we choose curves that fit the form
\begin{eqnarray}
f(x)=\frac{\left(f_0-h_{\infty}\right)^2}{\left(f_0-h_{\infty}\right)-x\delta}+h_{\infty}\label{eq:curve}
\end{eqnarray}
where $f_0$ is the height of the curve at $x=0,$ $\delta$ is the slope of the curve at $x=0,$ and $h_{\infty}=\lim_{x\to\infty}f(x).$ The values used in this manuscript are different between the two and three dimensional versions. (See supplementary materials.)

For these parameters, we can track the smallest distance, $w,$ between the $b$ spheres and their reflected counterparts as seen in Figure~\ref{fig:gaps}. From this figure, we see a very clear power law and conclude that in the limit of infinitely large bridges, no unwanted additional contacts are created. This means that regardless of the value of $n$ we choose, there are no overlaps for our Dionysian packing subject to the chosen curves $\mathcal{C}.$ Because the length of our bridges increase with $n$ but the other spatial dimensions do not, this construction results in packings with a density that falls like $n^{1-d}.$

\begin{figure}[]
\includegraphics[width=0.475\textwidth]{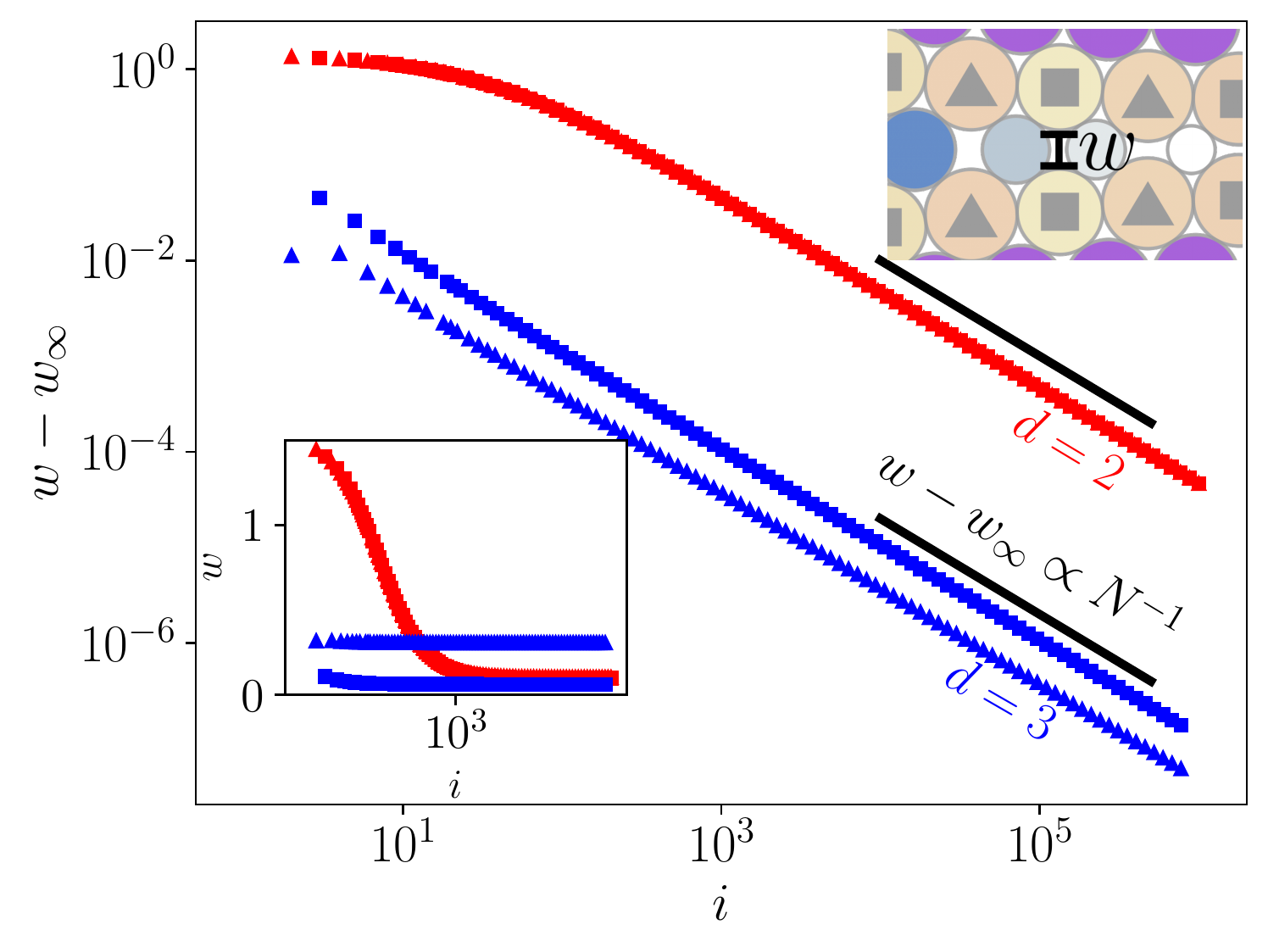}
\caption{Top right inset: demonstration of the definition of a gap for a circle. The $b$ circles, indexed by $i,$ oscillate in size and are separated into two categories labelled by squares and triangles. Bottom left inset: The gap value for both square and triangular marked spheres asymptotes in two and three dimensions. When the asymptotic gap value is subtracted, the gap sizes follow a power law of $N^{-1}$ as they reach their respective asymptotic values.}\label{fig:gaps}
\end{figure}

Using the aformentioned linear programming algorithm on our Dionysian packings, we find that they are both jammed and shear stable for every $n$ studied up to $n=105$ ($N=3145$) with packing fraction $0.0558$ in two dimensions and $n=25$ ($N=2731$) with packing fraction $0.0128$ in three dimensions.

In addition to demonstrating jamming and shear stability, we quantify the level of stability by calculating the shear, $G$, and bulk, $K$, moduli~\cite{askeland_science_2005, beer_mechanics_2009} shown in Figure~\ref{fig:moduli}. The two dimensional dionysian packing is isotropic and has a single shear modulus, $G.$ However, the three dimensional Dionysian packing, like the FCC crystal upon which it was based, has two independent shear moduli, $G_{100}$ and $G_{110}$\cite{ballato_poissons_1996}. These moduli in two dimensions can be calculated from the stiffness matrix as $K=\left(C_{11}+C_{12}\right)/2$ and $G=C_{33}.$ In three dimensions, these are calculated as $K=\left(C_{11}+2C_{12}\right)/2,$ $G_{100}=C_{44},$ and $G_{110}=\left(C_{11}-C_{12}\right)/2.$

\begin{figure}[]
\includegraphics[width=0.475\textwidth]{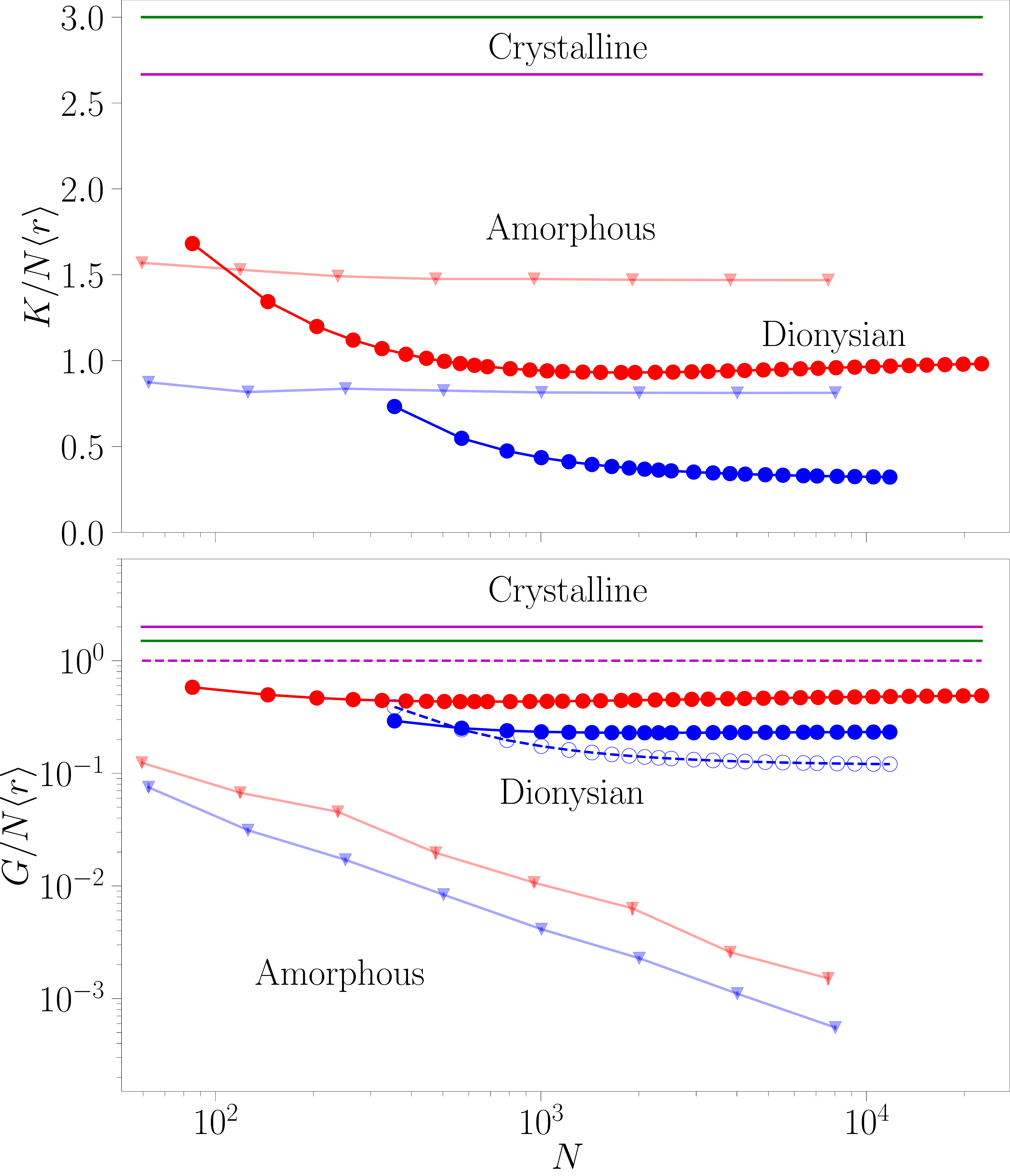}
\caption{The dimensionless bulk, $K,$ and shear, $G,$ moduli per sphere for Dionysian and amorphous packings in a unit cell as a function of the number of spheres, $N.$ The green line represents a two dimensional triangular packing, the magenta line represents a three dimensional FCC packing, and red and blue represent two dimensional and three dimensional packings respectively. The dashed curves with open symbols represent $G_{110},$ the shear modulus in direction $(1,1,0),$ whereas the solid curves with closed symbols represent $G_{100}.$ The results are exact for the Dionysian packings and crystals. For the amorphous systems, sufficiently many systems were sampled to make the standard error bars smaller than the plot markers. In the limit of large $N,$ the bulk modulus per sphere asymptotes to a positive value in two and three dimensions for all of the systems. The shear modulus for crystals and Dionysian packings plateaus for large $N$ indicating that these remain very stiff. On the other hand, the amorphous packings have a shear modulus that decreases like $1/N$~\cite{dagois-bohy_soft-sphere_2012}.}\label{fig:moduli}
\end{figure}

To compare the mechanical properties of Dionysian packings with other purely compressive solids, we also studied the properties of crystals and shear-stabilized jammed packings. We generated shear stabilized amorphous systems with monodisperse radii in three dimensions and $25\%$ polydispersity in two dimensions drawn from a log-normal distribution. We then used a modified FIRE algorithm~\cite{bitzek_structural_2006} that performs a constrained minimization with respect to both volume-preserving strains and positions as implemented in the pyCudaPacking software~\cite{morse_geometric_2014, charbonneau_universal_2016, morse_echoes_2017}. We created critically jammed and shear-stabilized packings by alternating between shear-stabilizing packings and uniformly decreasing the packing fraction and by extension the system pressure~\cite{charbonneau_jamming_2015}.

Figure~\ref{fig:moduli} demonstrates that crystals, shear-stabilized jammed systems, and Dionysian packings all have a bulk modulus per particle that plateaus to a fixed value in the limit of large $N.$ Similarly, the shear moduli per particle for crystals and Dionysian packings plateau for large $N.$ In contrast, we confirm the claim by Dagois-Bohy et al.~\cite{dagois-bohy_soft-sphere_2012} that the shear modulus in shear-stabilized jammed systems decreases like $1/N.$ These results indicate that Dionysian packings maintain their stability even as the density approaches zero whereas amorphous systems are only marginally stable in the thermodynamic limit. Remarkably, Dionysian packings can be created without sacrificing stiffness.

Extension of our procedure to higher dimensions can be proven to not be viable due to unavoidable overlapping of spheres (see supplementary materials). We conjecture that higher dimensional Dionysian packings also have arbitrarily low densities, but demonstrating this will require a novel construction.

\textit{Conclusions --} We find that the lower bound on density for mechanical stability of purely repulsive spheres is $0$ (Dionysian) and the upper bound is $1$ (Apollonian) in two and three dimensional sphere packings. In addition to this solution and the extension of our understanding of the limits associated with the jamming energy landscape, this discovery has implications for our fundamental understanding of mechanical stability. Where Apollonian packings can be used to create structures which fill space entirely, Dionysian packings can be used to create structures that utilize very little material and remain stiff. We prove that appreciably lighter weight materials can be constructed and give a road map for building them.

\textit{Acknowledgments --} We thank Aileen Carroll-Godfrey, Sean Ridout, James Sartor, and Mike Thorpe for helpful discussions and feedback. This work was supported by National Science Foundation (NSF) Career Award DMR-1255370 and the Simons Foundation No. 454939.

\bibliography{dionysus}

\clearpage
\widetext


\title{Supplementary Material for: Dionysian Hard Sphere Packings are Mechanically Stable at Vanishingly Low Densities}

\author{R. C. Dennis}
\author{E. I. Corwin}
\affiliation{Department of Physics and Materials Science Institute, University of Oregon, Eugene, Oregon 97403, USA.}

\maketitle

\section{Constructing a Dionysian Packing in Two Dimensions}
To construct a Dionysian packing in two dimensions, we do the following:
\begin{enumerate}
\item Create a chain of $n$ kissing circles labeled $a_1-a_n$ which have unit radius and centers that lie on a convex function $f(x)$ given by equation~6 such that the coordinates of each circle are $(x,f(x)).$ The values used in this manuscript can be found in table~\ref{tab:params}. If we give the bridges $h_{\infty}=1+\sqrt{3},$ we end up with nice monodisperse crystalline structures at infinity. However, because the radii of $b$ circles oscillate between two values, these values of $h_{\infty}$ will eventually cause overlapping to occur. To prevent this, we perturb these values by $0.05$
\item Place a circle $b_1$ of radius 1 that kisses $a_1$ and $a_2$
\item Place circle $b_m,$ where $m\in [2,n-1],$ such that it kisses $a_m,$ $a_{m+1},$ and $b_{m-1}$
\item Place circle $b_n$ such that it kisses $a_{n-1}$ and $a_n$ and so that its center lies at $a_{nx}+1$ where $a_{nx}$ is the $x$ coordinate of circle $a_n$
\item Place circle $c_m,$ where $m\in [1,n-1],$ such that it lies on $y=0$ and kisses circles $b_m$ and $b_{m+1}$
\item Reflect the ensemble of circles about the $x$ axis
\item Reflect the ensemble of circles about the line normal to the $x$ axis that passes through the center of $b_n$
\item Generate three of these bridges and connect them such that they share $a_1$ circles and lie along the contact vectors of the triangular packing
\item Contain the circle ensemble in a rhombus with periodic boundary conditions
\item Place seven identical circles inside the cavity between bridges such that they form a honeycomb pattern and each of the six outer circles touch two copies of $b_1$
\item Place six identical circles in the cavity each of which touches an $a$ circle and two of the circles in the honeycomb arrangement
\end{enumerate}
\section{Constructing a Dionysian Packing in Three Dimensions}
The construction process is very similar in three dimensions, but with the following changes
\begin{enumerate}
\item The values for the curve are different and can be found in table~\ref{tab:params}
\item The coordinates of $a$ spheres have the form $\begin{pmatrix}
a_x,a_y,0
\end{pmatrix},$ the coordinates of $b$ spheres have the form $\begin{pmatrix}
b_x,b_y,b_y
\end{pmatrix},$ and the coordinates of $c$ spheres have the form $\begin{pmatrix}
c_x,0,0
\end{pmatrix}$
\item The $a$ and $b$ spheres each have three copies that are rotated $45$ degrees about the $x$ axis
\item The sphere ensemble is reflected about the plane perpendicular to the $x$ axis that passes through the center of $b_n$
\item Six of these bridges are created and connected such that they share sphere $a_1$ and lie along the contact vectors of the primitive cell for the FCC packing
\item The spheres in the empty cavity formed by the bridges are different. Generate thirteen equal sized spheres, $f,$ in the shape of an fcc crystal such that one sphere is in the very middle of the cavity and the other twelve touch four $b$ spheres associated with the ends of the bridges. Connecting these bridges will naturally create two differently sized holes. In the six larger holes, create a dimer of equally sized circles, $m,$ such that they touch: each other, a $b$ sphere, two $a$ spheres, and an $f$ sphere. Also in these larger holes, place a sphere, $p,$ that touches eight of these $m$ sphere and an $f$ sphere. In the eight smaller holes, place a triangle of equally sized spheres, $q,$ that touch each other, three $a$ spheres and an $f$ sphere
\end{enumerate}
\begin{table}[h!]

\begin{tabular*}{\columnwidth}{@{\extracolsep{\stretch{1}}}  |c ||c| c| c|  }
  \hline
   $d$ & $f_{0}$ & $h_{\infty}$ & $\delta$ \\
  \hline\hline
  2 & $2\sqrt{3}$ & $\left(1+\sqrt{3}\right)+0.05$ & 0.01  \\
  \hline
  3 & $\sqrt{6}$ & $\left(1+\sqrt{2}\right)+0.025$ & 0.01  \\
  \hline
                          
\end{tabular*}
\caption{The values we used to parameterize curve $\mathcal{C}$ for various dimensions $d$ according to equation~9}
\label{tab:params}
\end{table}
For a visual representation of the construction in two dimensions, see FIG.~1.
\section{Trivial Extension to Higher Dimensions}
We can prove that extending this construction to higher dimensions will not work. The generalized construction is given by parameterizing the positions of the $a$ spheres as \begin{align*}
\vec{a}_{m}=\begin{pmatrix}
a_{mx},a_{my},0,0,\ldots
\end{pmatrix}~\textrm{and}~\vec{b}_{m}=\begin{pmatrix}
b_{mx},b_{my},b_{my},b_{my},\ldots
\end{pmatrix}
\end{align*}
for $m\in [1,n].$ The $a$ spheres will each have $2(d-1)$ copies given by
\begin{align*}
\begin{pmatrix}
a_{mx},-a_{my},0,0,\ldots
\end{pmatrix},
\begin{pmatrix}
a_{mx},0,a_{my},0,\ldots
\end{pmatrix},
\begin{pmatrix}
a_{mx},0,-a_{my},0,\ldots
\end{pmatrix},
\begin{pmatrix}
a_{mx},0,0,a_{my},0,\ldots
\end{pmatrix},\ldots
\end{align*}
and the $b$ spheres will each have $2^{d-1}$ copies given by
\begin{align*}
\begin{pmatrix}
b_{mx},-b_{my},b_{my},b_{my},\ldots
\end{pmatrix},
\begin{pmatrix}
b_{mx},b_{my},-b_{my},b_{my},\ldots
\end{pmatrix},
\begin{pmatrix}
b_{mx},-b_{my},-b_{my},b_{my},\ldots
\end{pmatrix},\ldots
\end{align*}

We consider $\vec{a}_1=\begin{pmatrix}
0,a_y,0,\ldots
\end{pmatrix}$ with unit radius and $\vec{b}_1=\begin{pmatrix}
1,b_y,b_y,\ldots
\end{pmatrix}$ with radius $b_r.$ If we enforce that these two spheres kiss, we can solve for $a_y.$ We can then find that $b_y$ has a maximum value of
\begin{align*}
b_y^*=\sqrt{\frac{b_r(b_r+2)}{(d-2)(d-1)}}.
\end{align*}
Because $b_1$ cannot overlap with one of it's copies, $b_y\geq b_r.$ This along with the above equation means that
\begin{align*}
b_r\leq \sqrt{\frac{b_r(b_r+2)}{(d-2)(d-1)}}
\end{align*}
or for $d>2,$
\begin{align*}
b_r\leq \frac{2}{d^2-3d+1}.
\end{align*}
We also know that in steady state, the sum of the radii for $b_m$ and $b_{m+1}$ will be $2.$ This means that setting $b_m$ to have a radius less than $1$ gives $b_{m+1}$ a radius greater than $1.$ Therefore, if we substitute $b_r=1,$ we arrive at an upper bound for $d:$
\begin{align*}
d\leq\frac{3+\sqrt{13}}{2}\approx 3.30278
\end{align*}
which means that this construction does not extend to dimensions higher than three. We do conjecture that a different construction procedure exists to generate Dionysian packings in higher dimensions.

\section{Minimal Curvature for three dimensional Dionysian Packings}
We remarked in the text that the curves $\mathcal{C}$ have a very subtle amount of curvature in three dimensions. Given $a$ and $b$ spheres of radius $1,$ the tightest Dionysian bridge configuration one can achieve has an $a$ sphere with $a_y=(1+\sqrt{2}).$ Any tighter and the $b$ spheres will overlap. The loosest configuration has $a_y=\sqrt{6}.$ Any looser and the $b$ spheres will no longer be contained. (See table~\ref{tab:params}). If our packing begins with the loosest configuration and ends with the tightest, the curve will decrease in height by $\sqrt{6}/(1+\sqrt{2})-1=1.46\%$ which is subtle.

\section{Amorphous Shear Stabilized Systems}
We generate amorphous shear stabilized systems by finding the traceless forces on strain degrees of freedom as given in equation~4 of the manuscript. We then use the FIRE algorithm on these strain degrees of freedom to adjust the lattice vectors and apply an affine strain. Because $\Tr(\eps)=0$ is just the linear approximation for volume conservation, we also rescale the lattice vectors after each minimization step. Once a shear stabilized packing is found, we alternate between minimizing the system and uniformly decreasing the radius of each particle in order to maintain the polydispersity. After rattlers are removed and the system is at one state of self stress, we find the mechanical properties.
\section{Computing the Stiffness Matrix}
We first consider our extended rigidity matrix for which
\begin{align}
R_{x\langle ij\rangle\left(k\gamma\right)} &= \left(\delta_{jk}-\delta_{ik}\right)n_{ij}^{\gamma} \\
R_{\eps\langle ij\rangle\left(\alpha\beta\right)}&=n_{ij}^{\alpha}n_{ij}^{\beta}\sigma_{ij}
\end{align}
for contact $\langle ij \rangle,$ particle $k,$ and dimension $\gamma.$ Here, also note that $n_{ij}^\gamma$ is the normalized contact vector between particle $j$ and particle $i$ and $\sigma_{ij}$ is the sum of the radii of particles $i$ and $j.$

In order to find the stiffness matrix, we define the extended hessian, which is
\begin{align}
H&=\begin{pmatrix}
H_{xx} & H_{x\eps} \\ H_{x\eps}^T & H_{\eps \eps}
\end{pmatrix}
\end{align}
where $H_{xx}$ is the second derivative of the energy function with respect to positional degrees of freedom, $H_{\eps\eps}$ is the second derivative with respect to strain degrees of freedom, and $H_{x\eps}$ are mixed derivatives.

From Hooke's law, we know that
\begin{align}
H\begin{pmatrix}
\Delta \vec{x} \\ \vec{\eps}
\end{pmatrix}=\begin{pmatrix}
-\vec{F} \\ \vec{\sigma}
\end{pmatrix}
\end{align}
where $\Delta \vec{x}$ is a perturbation vector of the particles and $\vec{\sigma}$ is the stress.
To find the stiffness matrix, we solve for the non-affine perturbation $\Delta \vec{x}_{\textrm{na}}$ that leave the spatial forces unchanged but imposes a stress:
\begin{align}
H\begin{pmatrix}
\Delta \vec{x}_{\textrm{na}} \\ \vec{\eps}
\end{pmatrix}=\begin{pmatrix}
\vec{0} \\ \vec{\sigma}
\end{pmatrix}.
\end{align}
If we solve this system of equations for $\vec{\sigma},$ we find that $C\vec{\eps}=\vec{\sigma}$ where the stiffness matrix is
\begin{align}
C=\left[H_{\eps\eps}-H_{x\eps}^T\left(H_{xx}\right)^{-1}H_{x\eps}\right].
\end{align}
The term, $\left(H_{xx}\right)^{-1}$ is the Moore-Penrose pseudoinverse~\cite{ben-israel_existence_2003} of the singular matrix $H_{xx}.$
While the algebra is simple, care must be taken to prove that it is valid to use the pseudoinverse for hyperstatic jammed packings. 

We can take this result further by considering that for systems without prestresses, such as ours, the extended hessian can also be written as
\begin{align}
H&=R^TR\\
&=\begin{pmatrix}
R_x^TR_x & R_x^TR_\eps \\
R_\eps^TR_x & R_\eps^TR_\eps
\end{pmatrix}
\end{align}
so that 
\begin{align}
C=\left[R_\eps^T R_\eps-R_\eps^T R_x\left(R_x^T R_x\right)^{-1}R_x^TR_{\eps}\right].\label{eq:hessianStiffness}
\end{align}
This can be further simplified by applying the singular value decomposition~\cite{klema_singular_1980} for $R_x.$ We can define the left singular vectors as $U$ which correspond to the linearly independent basis of bond stresses, the right singular vectors, $V,$ which correspond to normal modes, and $\Sigma$ which is the rectangular diagonal matrix of singular values. Given this,
\begin{align}
R_x=U\Sigma V^T.
\end{align}
If we make this substitution in equation~\ref{eq:hessianStiffness}, we find that
\begin{align}
C&=\left[R_\eps^T R_\eps-R_\eps^T U\Sigma\left(\Sigma^T\Sigma\right)^{-1}\Sigma^T U^TR_{\eps}\right]\\
&=R_{\eps}^T\left(\mathbf{1} - U\Sigma\left(\Sigma^T\Sigma\right)^{-1}\Sigma^T U^T\right)R_{\eps}\\
&=R_{\eps}^T\left(UU^T - U\Sigma\left(\Sigma^T\Sigma\right)^{-1}\Sigma^T U^T\right)R_{\eps}\\
&=R_{\eps}^TU\left(\mathbf{1} - \Sigma\left(\Sigma^T\Sigma\right)^{-1}\Sigma^T\right)U^TR_{\eps}.\label{eq:cexact}
\end{align}
The pseudoinverse of a diagonal matrix such as $\Sigma^T\Sigma$ is a diagonal matrix where the nonzero entries are inverted and the zero entries remain zero. To simplify this, we can rewrite $\Sigma.$ If we let there be $f$ floppy modes, $s$ states of self stress, and $z$ nonzero singular values, then we can choose to express $\Sigma$ as
\begin{align}
\Sigma = \begin{pmatrix}
Q_{z\times z} & \mathbf{0}_{z\times f} \\ \mathbf{0}_{s\times z} & \mathbf{0}_{s\times f}
\end{pmatrix}
\end{align} 
where $Q$ is the diagonal matrix of non-zero singular values and where we have explicitly assumed that $s>f.$ This assumption will always hold for shear stabilized packings where $f=d$ corresponds to trivial floppy modes. Also note that this form of $\Sigma$ assumes that the left and right singular vectors are arranged in a corresponding way. Substituting this equation into equation~\ref{eq:cexact}, we find that
\begin{align}
C&=R_{\eps}^TU\left(\mathbf{1} - \begin{pmatrix}
Q_{z\times z} & \mathbf{0}_{z\times f} \\ \mathbf{0}_{s\times z} & \mathbf{0}_{s\times f}
\end{pmatrix}\begin{pmatrix}
\left(Q_{z\times z}^2\right)^{-1} & \mathbf{0}_{z\times f} \\ \mathbf{0}_{f\times z} & \mathbf{0}_{f\times f}
\end{pmatrix}\begin{pmatrix}
Q_{z\times z} & \mathbf{0}_{z\times s} \\ \mathbf{0}_{f\times z} & \mathbf{0}_{f\times s}
\end{pmatrix}\right)U^TR_{\eps}\\
&=R_{\eps}^TU\left(\mathbf{1} - \begin{pmatrix}
\mathbf{1}_{z\times z} & \mathbf{0}_{z\times s} \\ \mathbf{0}_{s\times z} & \mathbf{0}_{s\times s}
\end{pmatrix} \right)U^TR_{\eps}\\
&=R_{\eps}^TU\begin{pmatrix}
\mathbf{0}_{z\times z} & \mathbf{0}_{z\times s} \\ \mathbf{0}_{s\times z} & \mathbf{1}_{s\times s}
\end{pmatrix}U^TR_{\eps}.
\end{align}
In this equation, the $\mathbf{1}_{s\times s}$ term corresponds to the entries associated with states of self stress. As such, 
\begin{align}
C=R_{\eps}^TSS^TR_{\eps}
\end{align}
where $S$ is the matrix of states of self stress for $R_x.$ Again, this expression is valid under the assumption that the packing has no prestress and is strictly jammed.

We can understand this result by considering how each term interacts with an arbitrary strain, $\vec{\eps}.$ This arbitrary strain results in bond stresses, $\vec{b}=R_\eps\vec{\eps}.$ However, these bond stresses will very likely result in perturbations of the particles which will bring the packing out of force balance. The $SS^T$ term removes any components of the bond stresses that are inconsistent with the states of self stress and therefore would cause particle movements. This new set of bond stresses is then passed through $R_\eps^T$ and gives the stress vector, $\vec{\sigma}.$

\end{document}